\documentclass[12pt]{article}
\usepackage{graphicx,amsfonts,amssymb,color,mathrsfs, amsmath,amsthm,bm}
\usepackage[margin=1in]{geometry}
\usepackage{indentfirst}
\usepackage{dsfont}
\usepackage{hyperref}
\usepackage{bookmark}
\usepackage{verbatim}
\usepackage{comment}
\usepackage{multirow}
\usepackage[round]{natbib}
\usepackage{bbm}
\usepackage{subfig}
\usepackage{float}
\usepackage{setspace}
\usepackage{titling}
\newcommand{\bs}{\mathbf{s}}

\newcommand{\bz}{\mathbf{z}}

\newcommand{\bL}{\mathbf{L}}

\newcommand{\bU}{\mathbf{U}}

\newcommand{\bR}{\mathbf{R}}

\newcommand{\bfmu}{\bm{\mu}}

\newcommand{\bftheta}{\bm{\theta}}

\newcommand{\bfeta}{\bm{\eta}}

\newcommand{\bfSigma}{\bm{\Sigma}}

\newcommand{\GP}{\mathcal{GP}}

\newcommand{\order}{\mathcal{O}}
\newcommand{\normal}{\mathcal{N}}

\newcommand{\domain}{\mathcal{D}}

\title{A Neural-Mean Vecchia Gaussian Process for Unified Argo Modeling}
\author{
Nian Liu \thanks{Department of Mathematics, University of Houston, Email: nliu9@central.uh.edu }
\and
Jian Cao \thanks{Department of Mathematics, University of Houston, Email: jcao21@central.uh.edu}
}
\date{}

\begin{document}
\maketitle
\vspace{-2cm}
\begin{abstract}
    Argo is an international program that collects temperature and salinity observations in the upper two kilometers of the global ocean. Most existing approaches for modeling Argo temperature rely on localized modeling within moving windows, first estimating a prescribed mean structure and then fitting Gaussian processes (GPs) to the mean-subtracted anomalies. Such strategies introduce challenges in designing suitable mean structures and defining local moving windows, often resulting in case-specific modeling choices. In this work, we propose a one-stop Gaussian process regression framework with a flexible mean structure and a generic spatio-temporal covariance function to jointly model Argo temperature data across broad spatial domains. Our fully data-driven approach achieves predictive performance that compares favorably with the established benchmarks that require moving-window regression and separate parametric mean estimation. To ensure scalability over large spatial regions, we employ the Vecchia approximation, which reduces the computational complexity from cubic to quasi-linear in the number of observations while preserving predictive accuracy. Using Argo data from January to March over the years 2007–2016, the same dataset used in prior benchmark studies, we demonstrate that our approach provides a unified and data-driven alternative for large-scale oceanographic analysis.

    \vspace{1em}
\end{abstract}

\section{Introduction}\label{Sec_Intro}

The international Argo program \citep{argo2000} provides global subsurface ocean observations using autonomous floats that drift with ocean currents. Each float measures temperature, pressure, and salinity as it ascends from approximately 2 km depth to the surface. In this paper, pressure and depth are used interchangeably, as a pressure of 1 decibar (dbar) roughly corresponds to a depth of 1 meter. A complete vertical set of measurements from a single float within a short time window is referred to as a profile, which approximately corresponds to one horizontal location (longitude and latitude). An observation denotes a single measurement (e.g., temperature or salinity) at a specific depth within a profile. Every ten days, floats transmit their profile data via satellite to the global Argo database. The resulting dataset covers vast spatial and temporal scales, offering rich information for ocean state estimation and prediction. However, effective analysis of these data requires spatio-temporal modeling that accounts for complex correlations across longitude, latitude, depth, and time—posing significant methodological and computational challenges.

A large body of work has addressed the modeling of temperature and salinity in Argo data. Most approaches adopt a two-stage modeling scheme, consisting of a parametric mean-field estimation followed by stochastic modeling of the residual (anomaly) field. In the first stage, a mean structure is typically fitted via weighted least squares; in the second, a Gaussian process (GP) or similar stochastic model captures the residual spatial-temporal variability. Representative works include \citet{RoemmichGilson09}, \citet{KuuselaStein18}, \citet{Hu2024}, and \citet{yargersh22}. \citet{RoemmichGilson09} divided the depth range into bins and locally fitted monthly mean fields using weighted least squares, followed by objective analysis \citep{Bretherton1976, Roemmich1983} of the anomalies. \citet{KuuselaStein18} interpolated temperatures onto three fixed pressure levels, estimated the Roemmich–Gilson (RG) mean field, and modeled anomalies using moving-window GP regression with a spatio-temporal covariance. \citet{Hu2024} used unweighted local windows to fit the RG mean field before applying the same GP model as \citet{KuuselaStein18}. \citet{yargersh22} instead estimated functional mean fields and introduced a spatio-temporal functional kriging model to jointly predict temperature and salinity as functions of depth. These two-stage approaches, while effective, can be suboptimal relative to a single-stage maximum likelihood estimation (MLE) when computationally feasible. Moreover, their uncertainty quantification applies only to the residual field rather than the original spatio-temporal response. Finally, many existing methods rely on localized modeling within moving spatio-temporal windows to achieve computational feasibility, which may introduce sensitivity to the design of moving windows.

Several other studies employed iterative optimization for Argo data modeling, which is generally less efficient than direct likelihood-based methods. For instance, \citet{GrayRiser15} decomposed an Argo variable into large-scale, small-scale, and noise components and iteratively estimated these using generalized least squares. \citet{Korte-Stapff24} jointly modeled temperature and salinity as a sum of a parametric mean and basis-function expansions with latent coefficients, estimated via an expectation–maximization (EM) algorithm. \cite{ParkKGG23} modeled ocean heat transport as the sum of a parametric mean structure with polynomial and trigonometric terms and a locally stationary GP with nugget effect, and estimated the mean and covariance parameters with an iterative EM algorithm. Although flexible, such iterative procedures are typically not scalable with the number of measurements used for training. Multivariate extensions have also been proposed, often using multivariate GPs. For example, \citet{BolinW2020} and \citet{YargerStoevHsing23} respectively developed SPDE-based and spectral-based multivariate Mat\'ern models to jointly model temperature anomalies at different depths, while \citet{MarySCJ25} introduced multivariate nonstationary GPs based on differential operators. \citet{YargerBhadra23} modeled temperature, salinity, and oxygen in the Southern Ocean by locally estimating a spatially varying mean and subsequently applying a stationary Gaussian model to the anomalies.

In this paper, we propose a one-stop Gaussian process regression for modeling Argo variables using a unified and data-driven approach. Motivated by the prevalence of two-stage anomaly-based modeling, our study revisits the question of whether a direct single-stage GP without a prescribed mean structure can achieve competitive performance. As flexible nonparametric models, GPs learn the underlying spatial-temporal patterns without requiring complex parametric means, especially when physical knowledge is unknown or incomplete. We demonstrate that a single-stage GP with commonly-used yet flexible mean and covariance structures performs favorably in comparison with two existing benchmarks, while avoiding case-specific mean and moving-window specifications. For empirical evaluation, we focus on the temperature field, the most widely studied Argo variable in the literature.

A major barrier to single-stage GP modeling is computational scalability. For $n$ observations, standard GP inference requires the Cholesky factorization of an $n \times n$ covariance matrix, which incurs $\order(n^3)$ floating-point operations, prohibitive for large-scale datasets such as Argo. Numerous scalable approximations have been developed, including covariance tapering \citep{Furrer06,Kaufman08,dzm09}, composite likelihood methods \citep{Varin2011_CL,Bevilacqua2012_CL,Eidsvik2014_CL}, Gaussian Markov random field approximations \citep{rue2005gaussian,Lindgren11}, and multi-resolution approaches \citep{Nychka15,Katzfuss17}. Here we employ the Vecchia approximation \citep{vecchia_estimation_1988,katzfuss_general_2021}, which imposes an ordered conditional independence assumption yielding a sparse inverse Cholesky factor of the covariance matrix. Theoretical and empirical studies \citep{stein04,Datta16,Guinness2018,Finley19,SparseCho_2021,cao_scalable_2022-1} have shown that this approximation preserves predictive accuracy while reducing computational complexity to quasi-linear in $n$. \citet{Guinness2021} further derived efficient gradient and Fisher information computations under the Vecchia framework, implemented in the \texttt{R} package \texttt{GpGp}. Although \citet{Guinness2021} briefly illustrated this approach using a subset of Argo temperature data, they did not provide a direct comparison against two-stage methods. The present work fills this gap by systematically evaluating the one-stop GP regression with Vecchia approximation against two established approaches.

The remainder of the paper is organized as follows. Section~\ref{Sec_Review_KS_Y} reviews two representative benchmark methods. Section~\ref{Sec_Model} details the Gaussian process model and the Vecchia approximation. Section~\ref{Sec_Experiments} presents empirical comparisons using Argo temperature data. Section~\ref{Sec_Conclusion} concludes with discussions and future directions.

\section{Representative methods in the literature}\label{Sec_Review_KS_Y}

\citet{KuuselaStein18} and \citet{yargersh22} represent two prominent approaches in the literature for modeling the Argo dataset. Both methods first estimate local mean fields and then fit stationary Gaussian processes (GPs) to the mean-subtracted anomaly fields. To ensure computational feasibility given the large data volume, they define local moving windows by longitude, latitude, and time, and assume independence across local windows. While the use of overlapping windows may help maintain continuity in parameter estimation and predictions, it is still subject to the boundary effect and the level of discontinuity would depend on the `step size' between windows. In addition, the two methods employ distinct prescribed regression functions for the mean structures, which are not necessarily physically grounded and may become suboptimal to data-driven methods when the training data is sufficient. 

\cite{KuuselaStein18} structure the mean based on the Roemmich-Gilson (RG) function \citep{RoemmichGilson09,Ridgway2002}, which includes linear and quadratic functions of the spatial coordinates and trigonometric functions of the temporal coordinate. Specifically, the RG mean field is
\begin{align}
    \mu(l,L,p,d) = &\beta_0+\beta_1l+\beta_2L+\beta_3p +\beta_4l^2+\beta_5L^2+\beta_6p^2\notag\\
    &+\sum_{k=1}^6\gamma_k\sin\left(2\pi k\frac{d}{365.25}\right)+\sum_{k=1}^6\delta_k\cos\left(2\pi k\frac{d}{365.25}\right),
    \label{RG_MeanField}
\end{align}
where $\{\beta_k\}_{k=0}^6$, $\{\gamma_k\}_{k=1}^6$ and $\{\delta_k\}_{k=1}^6$ are parameters to be estimated, $l$ is latitude , $L$ is longitude, $p$ is pressure, and $d$ is the day of the year. \cite{RoemmichGilson09} divided the pressure range of 0-2000 dbar into $58$ bins and locally fitted a monthly mean field for each bin at each horizontal grid point using weighted least-squares based on $300$ nearest measurements. \cite{KuuselaStein18} adopt the strategy of \cite{RoemmichGilson09} and assign weights inversely proportional to the distance $d_{\text{RG}}$
\begin{equation}
    d_{\text{RG}}(\bs,\bs_g) = \sqrt{(l-l_g)^2+(L-L_g)^2+\frac{5\left(200\text{km}\cdot(p-p_g)\right)^2}{p^2+p_g^2}},
\end{equation}
where we use $\bs$ to denote the triplet of latitude, longitude, and depth, i.e., $\bs = (l, L, p)$. $\bs_g = (l_g, L_g, p_g)$ and $\bs$ are the triplets corresponding to a grid point (i.e., a reference location) and a measurement used in the estimation of the local mean function at the grid point, respectively. Three pressure levels are used in \cite{KuuselaStein18}, namely 10 dbar, 300 dbar, and 1500 dbar, which have depth bands of width 10 dbar, 20 dbar, and 100 dbar, respectively. At each depth level, the local mean structures described in Equation (\ref{RG_MeanField}) were estimated independently at each grid point using $300$ measurements, consisting of $100$ observations each from the target pressure bin and its two adjacent bins, and for a large 2D horizontal grid with a band-width of $1/6^\circ$. The mean-subtracted anomalies were then modeled using a moving-window Gaussian process (GP), where GPs were fitted independently within spatial windows of $20^\circ \times 20^\circ$ and temporal ranges of either one month or three months over the years 2007–2016. The spatial window moved with a resolution of $1^\circ \times 1^\circ$. Each locally fitted GP employed a spatio-temporal exponential covariance function with an automatic relevance determination (ARD) structure \citep{neal2012bayesian}:
 \begin{equation}
 \label{KS_expCovFun}
     C(\Delta l,\Delta L,\Delta d) = \sigma^2\exp\left(-\sqrt{\left(\frac{\Delta l}{r_l}\right)^2+\left(\frac{\Delta L}{r_L}\right)^2+\left(\frac{\Delta d}{r_d}\right)^2}\right),
 \end{equation}
where $\sigma^2>0$ is the marginal variance; $\Delta l$, $\Delta L$, and $\Delta d$ are the differences in latitude, longitude and time (days); $r_l$, $r_L$, and $r_d$ are separate ranges for longitude, latitude, and day of the year, respectively. Posterior inference is performed locally, based on the ordinary kriging within the moving window where the point for prediction is located. 

\cite{yargersh22} proposed a spatio-temporal functional kriging approach that jointly models the temperature and salinity in the Argo dataset across depth. Denote by $W_{i,j}$ an Argo variable (temperature or salinity) measured at the $j$-th depth in the $i$th profile, noticing that the full set of covariates indexing a measurement is $(l_i,L_i,p_{i,j},d_i,y_i)$, where $(l_i,L_i)$ is the profile-dependent longitude-latitude pair, $p_{i,j}$ is the measurement-specific depth, $d_i$ and $y_i$ are the profile-dependent temporal coordinates, namely, the day of the year and the year number, respectively. The two-stage modeling approach in \cite{yargersh22} is based on the decomposition:
\begin{equation}
    \label{Y22_model}
    W_{i,j} = \mu(l_i,L_i,p_{i,j},d_i,y_i)+\xi(l_i,L_i,d_i,y_i,p_{i,j})+\epsilon_{i,j},
\end{equation}
where $\mu$ is a deterministic mean function, $\xi$ is a zero-mean stochastic process, and $\epsilon_{i,j}$ is a white noise that models the measurement error. The mean field $\mu$ includes linear and quadratic terms of the spatio-temporal coordinates:
\begin{equation}
    \mu(l,L,p,d,y) = \beta_{0,y}(p)+\beta_1(p)l+\beta_2(p)L+\beta_3(p)l^2+\beta_4(p)L^2+\beta_5(p)lL+\beta_6(p)d + \beta_7(p) d^2,
    \label{Y22_MeanField}
\end{equation}
where $\beta_{0,y}$ and $\{\beta_k\}_{k=1}^7$ are smooth functions with the support over the whole depth range and finite second-order moments. For each horizontal grid point at a $1^\circ$ resolution, the mean function is estimated using profiles located within $900$ kilometers of the grid point, collected from January to March over the years 2007–2016. The coefficients $\beta_{0,y}$ and $\{\beta_k\}_{k=1}^7$ in the mean function (\ref{Y22_MeanField}) are obtained via weighted least squares, with weights determined by the spatio-temporal distances between the selected profiles and the target grid point.

To obtain the anomaly process $\xi(\cdot)$, each $W_{i,j}$ is demeaned by subtracting the value of the local mean function of its nearest grid point, evaluated at the corresponding pressure level $p_{i,j}$. The anomaly $\xi$ is then modeled as
\begin{equation}
\label{Y22_KLexpansion}
\xi(l_i, L_i, d_i, y_i, p) = \sum_{k=1}^{10} Z_k(l_i, L_i, d_i, y_i)\phi_k(p),
\end{equation}
where $\{Z_k\}_{k=1}^{10}$ are ten independent Gaussian processes with exponential covariance functions that define the coefficients of $\{\phi_k\}_{k=1}^{10}$, a set of deterministic orthonormal basis, each being a linear combination of a univariate B-spline basis. The anomaly process, including the variance of the noise terms $\{\epsilon_{i,j}\}$, is estimated at each grid point of a $1^\circ$ resolution horizontal grid and at each pressure level, using all measurements (across depth) of the profiles located within $1{,}100$ kilometers of the target grid point via likelihood-based estimation. Posterior inference at any given spatio-temporal point is then performed using the fitted local model corresponding to its nearest grid point and the subset of responses within $1{,}100$ kilometers of the grid point horizontally.

Both approaches rely on complex mean structures and require sequential estimation of the mean field and the anomaly field. This two-stage modeling decouples mean estimation from anomaly inference, consequently, uncertainty from the anomaly inference is not propagated to the original measurements. Furthermore, making inference based on the model estimated for the nearest grid point is sensitive to users' choice of grid density and the window size that the local model is based on. It also ignores long-range dependence and increases the risk of overfitting. These limitations motivate a unified and fully data-driven approach, which we introduce in the following section.

\section{Unified GP modeling for Argo Temperature}\label{Sec_Model}

In this section, we propose to apply a one-stop GP regression to the temperature variable in the Argo dataset to show that the one-stop GP regression can achieve comparable and in some cases, even better, results than representative methods in the literature that rely on prescribed mean structures, local-window specification, and two-stage modeling. While our method is applicable to other Argo variables, we focus on temperature as it is the most widely studied variable in the literature.

\subsection{GP regression under ARD covariance functions}\label{SubSec_classicGP}
Gaussian processes (GPs) are a family of nonparametric stochastic models defined by a mean function $\mu(\cdot)$ and a covariance function $C(\cdot, \cdot)$. Specifically, for a collection of $n$ points $\{\bs_{i} \in \domain\}_{i = 1}^{n}$, where $\domain$ is the $q$-dimensional domain over which a GP is defined, their corresponding response variables $\bz = [z_{1}, z_{2}, \ldots, z_{n}]^\top$ jointly follow a multivariate normal (MVN) distribution:
\begin{equation*}
    \bz \sim \normal(\bfmu, \bfSigma),\ \bfmu = [\mu(\bs_{1}), \ldots, \mu(\bs_{n})]^\top, \mbox{ and } \bfSigma_{i,j} = C(\bs_i, \bs_j).
\end{equation*}
In this paper, we adopt a neural-network-based mean function $\mu(\cdot) = m_{\bfeta}(\cdot)$, where $\bfeta$ denotes the collection of network parameters. Integrating data-driven neural-networks into Gaussian Process (GP) frameworks allows for flexible, nonlinear mean modeling while preserving the rigorous probabilistic predictions characteristic of GPs \citep{Sigrist22_GPBoosting, Simchoni23_DNN, Saha23_RF, Zhan25_NNGeo}. The neural network $m_{\bfeta}$ is a feed-forward network mapping from $\mathbb{R}^q$ to $\mathbb{R}$ with three hidden layers. The sizes of the hidden layers depend on the size of the training data, but in general, we keep them moderate to prevent overfitting. This construction provides a flexible nonlinear mean structure that can capture complex large-scale spatio-temporal trends in the response. For the stochastic component, we employ a Mat\'ern covariance function \citep{Stein1999} augmented with an Automatic Relevance Determination (ARD) structure, which leads to the covariance function:
\begin{equation}
    \label{MaternCov}
    C(\bs_i,\bs_j;\bftheta) = \sigma^22^{1-\nu}/\Gamma(\nu)(\|\bR^{-1}(\bs_i-\bs_j)\|)^{\nu}K_{\nu}(\|\bR^{-1}(\bs_i-\bs_j)\|),
\end{equation}
where $\bR = \mbox{diag}(r_1,r_2,\dots,r_q)$ is a $q\times q$ diagonal matrix, with $r_{i}^{-1}$ being the (positive) scaling factor of the $i$-th dimension of $\domain$; $K_\nu$ is the modified Bessel function of the second kind with smoothness $\nu$; and $\Gamma$ is the gamma function. Notice that $\|\cdot\|$ is the Euclidean norm as the Mat\'ern covariance function is not guaranteed positive-definite under other forms of distances. The ARD structure accounts for anisotropy by allowing separate lengthscales for the $q$ dimensions of the domain, thereby providing extra flexibility while controlling the risk of overfitting assuming $n \gg q$. 

To account for measurement noise in practice, we incorporate a nugget effect into the GP model:
\begin{equation}
\label{GP_model}
    W \sim \GP( m_{\bfeta}(\cdot),C(\bs_i,\bs_j;\bftheta)+ \tau^2 \mathbbm{1}_{\{\bs_i= \bs_j\}}),
\end{equation}
where $W$ is the variable of interest, $m_{\bfeta}$ is the neural mean, $C$ is the covariance function parameterized by $\bftheta$, $\tau^2>0$ is the nugget variance, and $\mathbbm{1}$ is the indicator function. To summarize, the GP is parameterized by the neural mean parameters $\bfeta$, the covariance parameters $\bftheta=(\sigma^2,r_1,r_2,\dots,r_q, \nu)$, and the nugget variance $\tau^2$.

The conventional method for estimating GP parameters is through MLE, where one needs to iteratively compute the log-likelihood of the GP realization over $n$ locations. Given $\bz \sim \normal(\bfmu, \bfSigma)$, the likelihood for $\bz$ is
\begin{equation}
    \label{classicGP_jointPDF}
    f(\bz) = \left((2\pi)^n|\bfSigma|\right)^{-1/2}\exp\left(-\frac{1}{2}(\bz-\bfmu)^\top\bfSigma^{-1}(\bz-\bfmu)\right),
\end{equation}
where $|\bfSigma|$ denotes the determinant of $\bfSigma$. In practice, the matrix-inverse and the determinant are computed through the Cholesky factorization of the $n\times n$ covariance matrix $\bfSigma$, which requires $\mathcal{O}(n^2)$ memory and $\mathcal{O}(n^3)$ time. Consequently, the conventional method for estimating GP is limited to realizations over a few thousands of locations, for which, many scalable approximations for the GP likelihood were proposed. In this paper, we use the Vecchia approximation introduced in the next section.

\subsection{The Vecchia approximation for Gaussian processes}\label{SubSec_Vecchia}

The Vecchia approximation \citep{vecchia_estimation_1988} decomposes an $n$-dimensional joint density into a product of univariate conditional densities and truncates the conditioning sets to sizes controlled by a tuning parameter $m$ that only grows polylogarithmically with $n$, hence achieving an overall quasi-linear complexity with respect to $n$. Specifically, for a random vector $\bz = [z_{1}, \ldots, z_{n}]^\top$:
\begin{equation}
\label{VecchiaDensity}
    f(\bz) = f(z_1)\prod_{i=2}^nf(z_i\mid\bz_{1:i-1})\approx f(z_1)\prod_{i=2}^nf(z_i\mid\bz_{c(i)}),
\end{equation}
where for each $i\in\{2,\dots,n\}$, the conditioning set $c(i)$ is a subset of $\{1,\dots,i-1\}$ and the cardinality of $c(i)$ is $\min(i - 1, m)$. Choosing $m$ between $30$ and $50$ has been shown (e.g. by \citet{katzfuss_scaled_2022, Datta16, cao_variational_2023}) to achieve sufficiently high accuracy for $n$ on the magnitudes of tens of thousands under commonly used spatial covariance kernels. The ordering and the choice of the conditioning sets are the two tuning factors of the Vecchia approximation. For improved accuracy, \citet{Guinness2018} suggested using maximin ordering for higher accuracy while \citet{katzfuss_scaled_2022} proposed to choose the conditioning sets as the $m$ nearest neighbors in the scaled domain, where the $k$-th dimension of $\domain$ is scaled by $r_{k}^{-1}$, both of which have established algorithms that have quasi-linear complexity with respect to $n$.

Technically, the Vecchia approximation is applicable to beyond MVN densities but here in this paper, we mainly consider the cases where $f(\bz)$ is a MVN density function, under which its Vecchia approximation, denoted by $\hat{f}(\bz)$, amounts to another MVN distribution with the same mean parameter but a slightly different covariance matrix, denoted by $\hat{\bfSigma}$. Denoting the Cholesky factor and the inverse Cholesky factor of $\hat{\bfSigma}$ by $\hat{\bL}$ and $\hat{\bU} = \hat{\bL}^{-\top}$, respectively, $\hat{\bU}$ is an upper triangular matrix whose non-zero off-diagonal entries in the $j$-th column are exactly indexed by $c(j)$. Therefore, $\hat{\bU}$ is a sparse matrix with no more than $nm$ non-zero entries and furthermore, $\hat{\bU}$ can be computed in $\order(nm^3)$ time with straightforward parallelization over $n$ \citep{katzfuss_general_2021}. The GP likelihood function under the Vecchia approximation can be represented as:
\begin{equation}
    \label{VecchiaDensity_TX}
    \widehat f(\bz) = (2\pi)^{-n/2}|\hat{\bU}|\exp\left(-\frac{1}{2}\|(\bz-\bfmu)^\top\hat{\bU}\|^2\right),
\end{equation}
where $|\cdot|$ denotes the determinant and $\|\cdot\|^2$ denotes the square of the $L^2$ norm.

The Vecchia approximation also sheds light on the inference at unobserved locations. Conventionally, the GP prediction at unobserved locations requires a one-time $\order(n^3)$ Cholesky factorization of $\bfSigma$, followed by matrix-vector multiplications of complexity $\order(n^2)$ for each new prediction. In the framework of the Vecchia approximation, one can make prediction for $z^*$, the measurement at a new location $\bs^*$, using only the $m$ nearest neighbors of $\bs^*$, essentially assuming $f(z^* | \bz) \approx f(z^* | \bz_{c^*})$, where $c^* \subset \{1, \ldots, n\}$ is the set of indices of the $m$ nearest neighbors of $\bs^*$, hence achieving an $\order(m^3)$ complexity for the prediction at each new location, with $m \ll n$.

In summary, the Vecchia approximation reduces the computational cost of GP regression and inference to quasi-linear complexity in the number of training measurements $n$, while allowing straightforward parallelization over $n$. We fully leverage these advantages in applying GP regression to the large-scale Argo dataset.

\subsection{Implementation for the Argo dataset}\label{SubSec_SSGP}

We propose to model an Argo variable of interest, for example, the temperature variable, using a one-stop GP regression, with the GP construction \eqref{GP_model} and the Vecchia likelihood approximation introduced in Section \ref{SubSec_Vecchia}. 

As in \cite{yargersh22}, we define the dimensions of the domain $\domain$ to include the latitude $l$, the longitude $L$, the pressure $p$, the year $y$, and the day of the year $d$. We include year as an input in the covariance structure to capture inter-annual dependence in ocean temperatures. Incorporating it within the ARD covariance offers a flexible, data-driven approach for determining the relevance/importance of covariates. At a given latitude, longitudes of $180^\circ$E ($L=180$) and $180^\circ$W ($L=-180$) correspond to the same physical location. Therefore, we represent the circular variable $L$ using $L_s=\sin(\pi L/180)$ and $L_c=\cos(\pi L/180)$. To account for seasonality and ensure continuity between December and January, we similarly transform $d$ into $d_s=\sin(2\pi d/365)$ and $d_c=\cos(2\pi d/365)$. Thus, each input vector is seven-dimensional, $\bs = (l,L_s,L_c,p,y,d_s,d_c)$. For benchmarking experiments restricted to a single month (aggregated over multiple years), we omit $d_s$ and $d_c$, assuming strong intra-month temporal correlation and favoring model parsimony. 

Our proposed one-stop model estimation jointly optimizes $\bfeta$, $\bftheta$, and $\tau^2$ via MLE, except for the smoothness parameter $\nu$. As in \cite{KuuselaStein18} and \cite{yargersh22}, we fix the smoothness parameter at $\nu=0.5$, which corresponds to the exponential covariance as a special case of the Mat\'ern class. The Argo dataset is collected by approximately $4{,}000$ floats, which together provide about $13{,}000$ profiles per month, each containing an average of $200$ measurements, resulting in over $30$ million measurements annually for each Argo variable. This massive data volume highlights the computation efficiency of the Vecchia approximation. We apply the proposed one-stop GP regression to the global Argo dataset and two regional subsets containing about $30{,}000$ profiles and compare its performance against two representative benchmarks from the literature. Our proposed method is implemented in a PyTorch-based package \citep{NEURIPS2019_9015}. Model parameters are estimated by optimizing the Vecchia-approximated log-likelihood using stochastic gradients and automatic differentiation. We set $m = 30$ for GP regression and $m = 100$ for the inference at testing locations.

\section{Numerical Comparison}\label{Sec_Experiments}

\subsection{Experiment setup}
\label{Sec_Experiments_setup}

In the following experiments, we use the Argo temperature data collected from January to March over the years from 2007 to 2016 to compare our proposed one-stop GP to the moving-window GP regression (MWGP) from \citet{KuuselaStein18} and the spatio-temporal kriging for functional data (KFD) from \citet{yargersh22}. We use the pre-screened Argo data as in \cite{KuuselaStein18}, where measurements deemed contaminated are removed. Both \citet{KuuselaStein18} and \cite{yargersh22} demonstrated their performances by predicting the Argo temperature in February, with the former proposing two models addressing using either the same month's (i.e., February's) or the whole three months' data for training, respectively. For a more comprehensive comparison, we also set up two training scenarios, using only February's or all three months' temperature measurements from the training profiles for model estimation, to predict February's temperature measurements of the testing profiles.

The two time windows, February alone and January through March, correspond to 73,534 and 234,401 profiles, respectively, over the years 2007–2016. We consider three distinct training–testing scenarios: one global and two regional. The global dataset includes floats distributed worldwide, while the two regional subsets each comprises floats within a 4,250-kilometer radius—one centered in the open Pacific Ocean and the other in the Atlantic Ocean, capturing more coastal observations. The floats in each scenario are then randomly split into 80\% for training and 20\% for testing. To ensure consistency with \citet{KuuselaStein18} and \citet{yargersh22}, we further downscale the testing profiles to those collected in Februaries. Besides the selection of profiles, \citet{KuuselaStein18} and \citet{yargersh22} both considered measurements of profiles localized to three pressure levels, namely 10, 300, and 1{,}500 dbar, covering the depth ranges of $[5,15]$ dbar, $[290,310]$ dbar, and $[1450, 1550]$ dbar, respectively. Following their approach, we also subset the measurements of the profiles within each training–testing scenario into these three depth bins, yielding nine datasets in total. In the global scenario, the datasets corresponding to the 10 dbar and 300 dbar pressure levels each contain approximately 600,000 training measurements and 50,000 testing measurements, while the dataset corresponding to 1,500 dbar contains approximately 1,500,000 training measurements and 120,000 testing measurements. In the regional scenarios, the datasets for the 10 dbar and 300 dbar levels contain approximately 85,000 training and 7,000 testing measurements, while the 1,500 dbar datasets contain approximately 220,000 training and 16,000 testing measurements. 

Both \citet{KuuselaStein18} and \citet{yargersh22} used localized methods in their estimations of the mean structure and the subsequent estimations of the anomaly field. However, the two methods differ in terms of their mean structure parameterizations, sizes of moving windows, and the resolution of their chosen horizontal grids; please refer to Section~\ref{Sec_Review_KS_Y} for more details. When implementing the MWGP method, we compute the RG mean field (\ref{RG_MeanField}) at each $1/6^\circ \times 1/6^\circ$ horizontal grid point for each month from January to March at the three pressure levels, using the whole three-month training dataset across depth, resulting in nine mean fields in total. MWGP further estimated local covariance parameters via MLE using training-profile anomalies within a $20^\circ \times 20^\circ$ moving window centered at each $1^\circ$ ocean grid point. \cite{KuuselaStein18} tested the MWGP method with two different covariance parameterizations, corresponding to their Models 2 and 5, respectively. Specifically, the former considers only spatial correlation, defining two range parameters $r_l$ and $r_L$ for longitude and latitude, respectively, while the latter also considered temporal correlation by including an extra range parameter $r_d$ for `day of the year'. The prediction for the testing profiles at the target pressure level was the sum of the mean predictions and the anomaly kriging results. The two models are denoted by MWGP-S and MWGP-ST, respectively. For the KFD method, the functional mean with depth-varying coefficients (\ref{Y22_MeanField}) is locally estimated at each $1^\circ \times 1^\circ$ (horizontal) grid point, using January–March training profiles located within 900 km of that point. Subsequently, analogous to the $20^\circ \times 20^\circ$ moving-window approach, the mean-subtracted training anomalies within 1,100 km of each grid point are used to locally estimate the covariance structures of the stochastic process $\xi$ and the functional nugget $\epsilon$ in Equation (\ref{Y22_model}). It is worth noting that our implementations of \citet{KuuselaStein18} and \citet{yargersh22} have minor differences from their original implementations. \citet{KuuselaStein18} and \citet{yargersh22} estimated the mean structures using data from both training and testing profiles, incorporating temperature measurements across all depths, before dividing the resulting anomalies into pressure bins for kriging. To align with our training-testing setup, we follow their procedure for mean estimation using measurements across depths but apply it to data from the training profiles only. Notice that due to the usage of weighted least square regression in \citet{KuuselaStein18} and \citet{yargersh22}, the fitted mean structures are different at the three pressure levels even using the same collection of training profiles.

Our proposed one-stop GP directly maximizes the GP log-likelihood under the Vecchia approximation over the training profiles for each of the nine training-testing scenarios, amounting to a unified modeling procedure. As introduced in Section~\ref{SubSec_SSGP}, our model parameters include $\bfeta$, $\tau^2$, and $\bftheta=(\sigma^2,r_1,r_2,\dots,r_7,\nu)$, where $r_1,r_2,\dots,r_7$ correspond to $(l,L_s,L_c,p,y,d_s,d_c)$, and $\nu$ is fixed at $0.5$. We refer to this model as Vecchia-based GP with 7-dimensional inputs (VGP-7D). When only profiles in Februaries are used for training, we modify the inputs to five dimensional vectors that omit $d_s$ and $d_c$, which is referred to as VGP-5D. The GPs trained with only the profiles in Februaries are used for comparison with the MWGP-S from \citet{KuuselaStein18} while those trained with all three months' profiles are compared against MWGP-ST as well as the KFD method.

The criteria we use to compare the posterior inference made by different methods are:
the root-mean-square error (RMSE), the third quartile of the absolute error (Q3AE), the median absolute error (MdAE), the mean absolute error (MAE), the continuous ranked probability score (CRPS), and the coefficient of determination ($R^2$). RMSE is the most commonly used performance metric while Q3AE, MdAE, and MAE were used in \citet{KuuselaStein18} and \citet{yargersh22} and hence are also included here. CRPS is defined as
\begin{equation}
    \label{CRPS}
    \text{CRPS} = \frac{1}{N}\sum_{p=1}^N\left(\int_{-\infty}^{W(\bs_p^*)}\left(\Phi_p^*(x)\right)^2\mathrm{d}x + \int_{W(\bs_p^*)}^\infty\left(1-\Phi_p^*(x)\right)^2\mathrm{d}x\right),
\end{equation}
where $W(\bs_p^*)$ is the observed temperature at prediction location $\bs_p^*$, $N$ is the number of measurements in the testing dataset, and $\Phi_p^*(\cdot)$ is the cumulative distribution function of the prediction distribution, which is typically assumed normal. CRPS measures the quality of both point prediction and uncertainty quantification while $R^2$ shows the prediction error relative to the total variation. Notice that the magnitudes of the Argo variables, including the temperature, vary substantially with depth and typically, the variances in depth bins closer to the surface are much higher, for which the $R^2$ is more comparable across depths.

\subsection{The Global scenario} 
\label{Sec_Experiments_glob}

\begin{figure}[htbp]
    \centering
    \caption{The number of training profiles within a $20^\circ\times20^\circ$ moving-window centered at each grid point of a $1^\circ$ resolution horizontal grid. Region I is centered at $(30^\circ \text{S}, 150^\circ \text{W})$, corresponding to Table~\ref{tab:prediction_compare_regional1}; Region II is centered at $(30^\circ \text{N}, 30^\circ \text{W})$, corresponding to Table~\ref{tab:prediction_compare_regional2}.}
    \subfloat[Training profile density during January to March]{\includegraphics[width=0.483\linewidth]{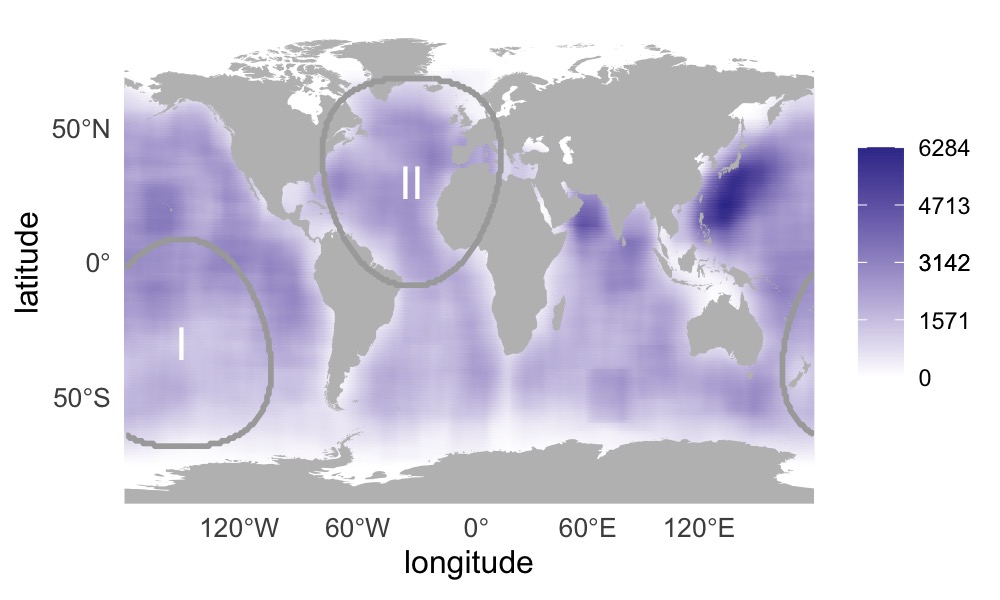}
    \label{fig:KSM5_nRes}
    }
      \hfill
    \subfloat[Training profile density during February]{\includegraphics[width=0.483\linewidth]{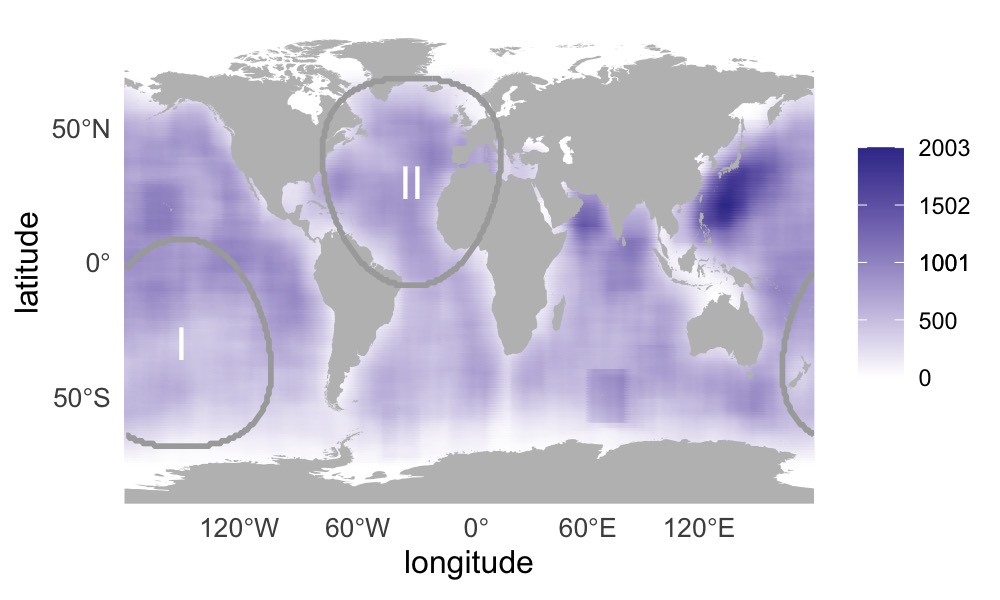}
    \label{fig:KSM2_nRes}
    }
\end{figure}

From the floats collected over the three-month period of January to March from 2007 to 2016, $80\%$ and $20\%$ are randomly assigned to the training and testing sets, respectively, with the testing profiles further restricted to those collected in February. 
Figure~\ref{fig:KSM5_nRes} presents the density of training profiles, measured by the number of profiles in each $20^\circ\times20^\circ$ moving-window, which is the same window size as used by \citet{KuuselaStein18} for modeling and predicting the anomaly field. The density distribution of training profiles in February presented in Figure \ref{fig:KSM2_nRes} follows the same pattern as that of the three-month training data, but differing by a factor $1/3$.

The predictive performance of the three methods, namely MWGP, KFD, and the proposed one-stop GP, is summarized in Table~\ref{tab:prediction_compare_global}. Overall, the one-stop GP performs well, outperforming the two benchmarks in four out of six cases, with only small gap in other cases. All methods yield great $R^2$ values, demonstrating their effectiveness for globally distributed Argo profiles. With the exception of the 1{,}500 dbar pressure level under the three-month training window, where KFD achieves the best Q$_3$AE, MdAE, and CRPS while VGP-7D yields the best RMSE, MAE, and $R^2$, the model rankings remain consistent across all evaluation criteria. This agreement across metrics indicates that the observed performance differences are consistent in point prediction accuracy and uncertainty quantification. The fitted covariance parameters of our proposed model are provided in the appendix. The fitted scale parameter in the one-stop GP model is larger at the 10 dbar and 300 dbar levels and smaller at 1,500 dbar, consistent with reduced temperature variability at greater depths. The fitted range parameters indicate that the dependence across latitude is relatively weak, whereas dependence across pressure and day of the year is strong. The estimated nugget variance is approximately 0.002 across all cases.

\begin{table}[t] 
    \centering
    \caption{Model performance comparison in predicting the testing temperature measurements at three pressure levels in the global scenario. The first column shows the month-depth pair, with depth having the unit of dbar. For each scenario, the best value of each performance metric is indicated in \textbf{\textit{bold}}. 
    }
    \begin{tabular}{c|c|c|c|c|c|c|c}
        \hline
        Data & Model & RMSE & Q$_3$AE & MdAE & MAE & $R^2$ &CRPS\\ \hline
        \multirow{2}{*}{Feb - 10} 
        & MWGP-S   &0.9755 & 0.9072 & 0.4649 & 0.6739 & 0.9869&0.5356  \\
        \cline{2-8} & VGP-5D & \textbf{\textit{0.7854}} & \textbf{\textit{0.6178}} & \textbf{\textit{0.3129}} & \textbf{\textit{0.4939}} & \textbf{\textit{0.9923}} & \textbf{\textit{0.3865}}\\ \hline
        \multirow{3}{*}{Jan-Mar - 10} & MWGP-ST  &0.7204 & 0.5566 & \textbf{\textit{0.2723}} & 0.4432 & 0.9928&0.3279\\ 
        \cline{2-8}
        & KFD   &\textbf{\textit{0.7054}} & \textbf{\textit{0.5541}} & 0.2805 & \textbf{\textit{0.4390}} & \textbf{\textit{0.9940}}&\textbf{\textit{0.3272}}\\ 
        \cline{2-8}
        & VGP-7D   & 0.7642 & 0.6078 & 0.2972 & 0.4746 & 0.9927 & 0.3693\\\hline
        \multirow{2}{*}{Feb - 300} 
        & MWGP-S   & 0.9241 & 0.7259 & 0.3519 & 0.5749 & 0.9589&0.4693\\ 
        \cline{2-8} & VGP-5D & \textbf{\textit{0.8491}} & \textbf{\textit{0.6594}} & \textbf{\textit{0.3127}} & \textbf{\textit{0.5255}} & \textbf{\textit{0.9679}} & \textbf{\textit{0.4227}}\\ \hline
        \multirow{3}{*}{Jan-Mar - 300} & MWGP-ST   &0.8500 & 0.6564 & 0.3114 & 0.5210 & 0.9644&0.3940\\
        \cline{2-8}
        & KFD   & \textbf{\textit{0.7891}} & \textbf{\textit{0.6175}} & \textbf{\textit{0.2977}} & \textbf{\textit{0.4936}} & \textbf{\textit{0.9723}}&\textbf{\textit{0.3705}}\\ 
        \cline{2-8}
        & VGP-7D   & 0.8271 & 0.6408 & 0.3052 & 0.5138 & 0.9695 & 0.4139\\\hline
        \multirow{2}{*}{Feb - 1500}
        & MWGP-S   &0.1519 & 0.1188 & 0.0609 & 0.0952 & 0.9870 & 0.0772\\ 
        \cline{2-8} & VGP-5D &\textbf{\textit{0.1231}} & \textbf{\textit{0.1085}} & \textbf{\textit{0.0582}} & \textbf{\textit{0.0834}} & \textbf{\textit{0.9886}} & \textbf{\textit{0.0648}}\\ \hline
        \multirow{3}{*}{Jan-Mar - 1500} & MWGP-ST   &0.1452 & 0.1197 & 0.0620 & 0.0938 & 0.9880&0.0683\\
        \cline{2-8}
        & KFD   & 0.1257 & \textbf{\textit{0.1078}} & \textbf{\textit{0.0553}} & 0.0835 & 0.9882 & \textbf{\textit{0.0642}}\\ 
        \cline{2-8}
        & VGP-7D   & \textbf{\textit{0.1205}} & 0.1079 & 0.0560 & \textbf{\textit{0.0817}} & \textbf{\textit{0.9891}} & 0.0643\\ \hline
    \end{tabular}
    \label{tab:prediction_compare_global}
\end{table}

\begin{figure}[ht]
    \centering
    \caption{Residuals at 1500 dbar in the global Jan-Mar scenario.}
    \label{fig:Res_1500}
    \subfloat[KFD]{
    \includegraphics[width=0.483\linewidth]{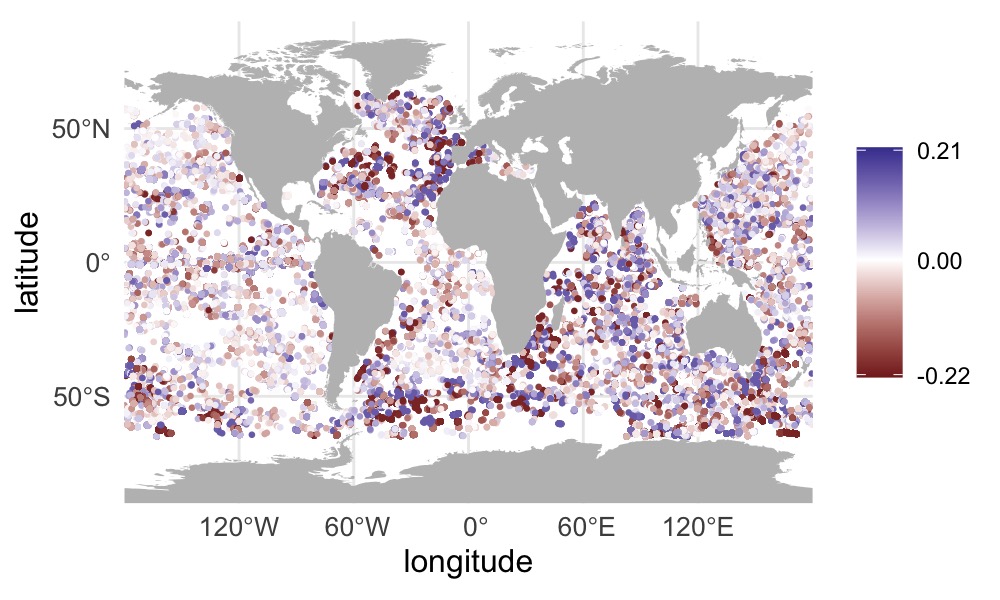}
    \label{fig:Res_1500_KFD}
    }
    \hfill
    \subfloat[VGP-7D]{
    \includegraphics[width=0.483\linewidth]{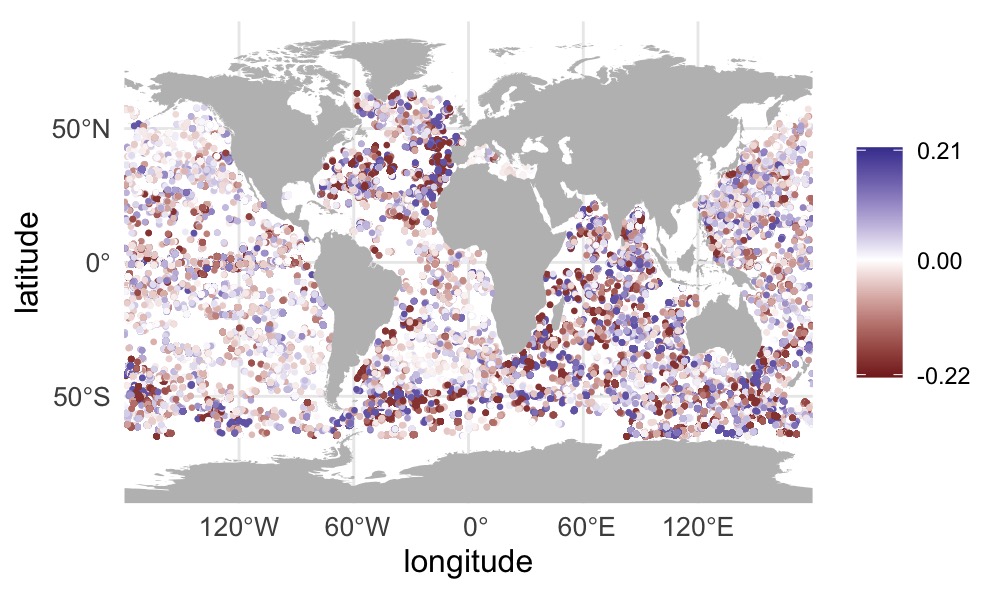}
    \label{fig:Res_1500_VGP-7D}
    }\\
    \subfloat[MWGP-ST]{
    \includegraphics[width=0.483\linewidth]{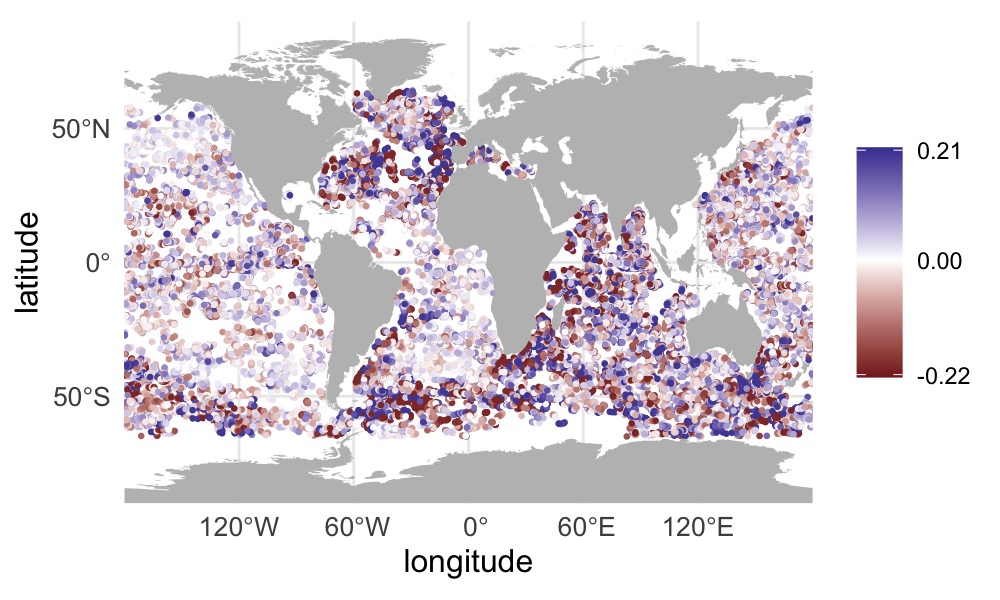}
    \label{fig:Res_1500_MWGP}
    }
\end{figure}

When the training profiles were limited to February, VGP-5D uniformly outperformed MWGP-S, highlighting the proposed model's capability in capturing the spatio-temporal dependence. Expanding the training window to three months improved the performance of both MWGP and VGP, reflecting their ability of assimilating the intra-year temporal dependence. Under the three-month scenarios, VGP remains competitive, outperforming both benchmarks in one out of three cases, while being close to the best performance in other cases. Figure \ref{fig:Res_1500} presents the prediction residuals for KFD, VGP, and MWGP at the pressure of 1,500 dbar. The spatial patterns are similar across all methods, with larger errors occurring near coastlines and along major ocean currents, indicating consistency between the proposed method and the two established benchmarks. In principle, the proposed one-stop GP method can jointly model measurements across depth, but in this paper, we do not pursue this route as the vertical structure of the Argo data may require more sophisticated nonstationarity kernels than the ARD Mat\'ern covariance considered in this paper \citep{RoemmichGilson09,Gasparin15,salvana20223d}.

\subsection{The Regional scenarios}
\label{Sec_Experiments_sub}

For the two regional training–testing experiments, we selected float subsets within 4,250 km of $(30^\circ \text{S}, 150^\circ \text{W})$ and $(30^\circ \text{N}, 30^\circ \text{W})$, respectively, resulting in approximately 30,000 profiles per region, about 10,000 of which were collected in February. The visualization of profile density over regions I and II is also provided in Figures~\ref{fig:KSM5_nRes}–\ref{fig:KSM2_nRes}. Shrinking the spatial domain for MWGP and KFD had little effect on their performances, as both rely on local model estimation and prediction based on smaller moving windows. The estimated covariance parameters for the one-stop GP are consistent with those observed in the global scenario. In particular, the scale parameter is significantly smaller at the depth level of 1500 dbar than at the other two depth levels and the correlation remains relatively weak across latitude.

The performance comparisons for the two regional scenarios are summarized in Tables~\ref{tab:prediction_compare_regional1} and~\ref{tab:prediction_compare_regional2}, respectively, where the differences across metrics between MWGP, KFD, and VGP are generally small. Overall, VGP outperformed both benchmarks in six out of the total twelve scenarios, with three scenarios each in one-month and three-month training-time windows. The performance gaps in the remaining scenarios are small. Based on $R^2$ and CRPS, prediction accuracy at 10 dbar and 300 dbar is higher in the region centered at $(30^\circ \text{S}, 150^\circ \text{W})$ than in the region centered at $(30^\circ \text{N}, 30^\circ \text{W})$, likely due to greater temperature variability in coastal regions at shallow depths. Notably, the one-stop GP performs better in the region centered at $(30^\circ \text{N}, 30^\circ \text{W})$, achieving the highest $R^2$ in five of six cases and the lowest CRPS in four cases. This is likely due to the neural mean, which more effectively captures the large-scale signal in temperature fields induced by strong currents such as the Gulf Stream.

\begin{table}[t]
    \centering
    \caption{Model performance comparison in predicting the temperature measurements at selected pressure levels from the testing profiles in the region centered at $(30^\circ \text{S}, 150^\circ \text{W})$. The first column shows the month-depth pair, with depth having the unit of dbar. For each scenario, the best value of every evaluation criterion is indicated in \textbf{\textit{bold}}.}
    \begin{tabular}{c|c|c|c|c|c|c|c}
        \hline
        Data & Model & RMSE & Q$_3$AE & MdAE & MAE & $R^2$ &CRPS\\ \hline
        \multirow{2}{*}{Feb - 10} 
        & MWGP-S   &\textbf{\textit{0.5402}} & \textbf{\textit{0.5472}} & 0.2993 & \textbf{\textit{0.3958}} & \textbf{\textit{0.9953}}& \textbf{\textit{0.2928}} \\ 
        \cline{2-8} & VGP-5D & 0.5631 & 0.5484 & \textbf{\textit{0.2884}} & 0.4028 & 0.9952 & 0.3011\\\hline
        \multirow{3}{*}{Jan-Mar - 10} & MWGP-ST  &\textbf{\textit{0.5004}} & 0.4757 & \textbf{\textit{0.2398}} & 0.3517 & 0.9959&\textbf{\textit{0.2546}}\\ 
        \cline{2-8}
        & KFD   &0.5019 & 0.4788 & 0.2562 & 0.3567 & \textbf{\textit{0.9964}}&0.2622\\ 
        \cline{2-8}
        & VGP-7D   &0.5029 & \textbf{\textit{0.4699}} & 0.2436 & \textbf{\textit{0.3505}} & 0.9962 & 0.2644\\\hline
        \multirow{2}{*}{Feb - 300} 
        & MWGP-S   &0.7250 & 0.7189 & 0.3838 & 0.5269 & 0.9666&0.4002\\ 
        \cline{2-8} & VGP-5D & \textbf{\textit{0.6839}} & \textbf{\textit{0.6638}} & \textbf{\textit{0.3375}} & \textbf{\textit{0.4854}} & \textbf{\textit{0.9721}} & \textbf{\textit{0.3655}}\\\hline
        \multirow{3}{*}{Jan-Mar - 300} & MWGP-ST   &0.7107 & 0.7078 & 0.3751 & 0.5181 & 0.9678&0.3763\\
        \cline{2-8}
        & KFD   & \textbf{\textit{0.6530}} & \textbf{\textit{0.6412}} & \textbf{\textit{0.3389}} & \textbf{\textit{0.4685}} & \textbf{\textit{0.9745}}&\textbf{\textit{0.3406}}\\ 
        \cline{2-8}
        & VGP-7D   &0.6755 & 0.6523 & 0.3437 & 0.4811 & 0.9728 & 0.3603\\\hline
        \multirow{2}{*}{Feb - 1500}
        & MWGP-S   & \textbf{\textit{0.1031}} & \textbf{\textit{0.0959}} & \textbf{\textit{0.0521}} & \textbf{\textit{0.0720}} & 0.9429&\textbf{\textit{0.0541}}\\ 
         \cline{2-8} & VGP-5D & 0.1118 & 0.1022 & 0.0539 & 0.0785 & \textbf{\textit{0.9486}} & 0.0604\\\hline
        \multirow{3}{*}{Jan-Mar - 1500} & MWGP-ST   &\textbf{\textit{0.1063}} & 0.0989 & \textbf{\textit{0.0515}} & 0.0735 & 0.9402&\textbf{\textit{0.0535}}\\
        \cline{2-8}
        & KFD   &0.1113 & \textbf{\textit{0.0955}} & 0.0517 & \textbf{\textit{0.0734}} & \textbf{\textit{0.9491}}&0.0568\\ 
        \cline{2-8}
        & VGP-7D   &0.1133 & 0.1015 & 0.0532 & 0.0783 & 0.9472 & 0.0600\\\hline
    \end{tabular}
    \label{tab:prediction_compare_regional1}
\end{table}

\begin{table}[t]
    \centering
    \caption{Model performance comparison in predicting the temperature measurements at selected pressure levels from the testing profiles in the region centered at $(30^\circ \text{N}, 30^\circ \text{W})$. The first column shows the month-depth pair, with depth having the unit of dbar. For each scenario, the best value of every evaluation criterion is indicated in \textbf{\textit{bold}}.}
    \begin{tabular}{c|c|c|c|c|c|c|c}
        \hline
        Data & Model & RMSE & Q$_3$AE & MdAE & MAE & $R^2$ &CRPS\\ \hline
        \multirow{2}{*}{Feb - 10} 
        & MWGP-S   &1.0137 & 0.6674 & 0.3480 & 0.5759 & 0.9818& 0.5287 \\ 
        \cline{2-8} & VGP-5D & \textbf{\textit{0.8290}} & \textbf{\textit{0.5181}} & \textbf{\textit{0.2664}} & \textbf{\textit{0.4512}} & \textbf{\textit{0.9888}} & \textbf{\textit{0.3729}}\\\hline
        \multirow{3}{*}{Jan-Mar - 10} & MWGP-ST  &0.8745 & 0.4865 & 0.2306 & 0.4494 & 0.9865&0.3456\\ 
        \cline{2-8}
        & KFD   &0.8760 & \textbf{\textit{0.4305}} & \textbf{\textit{0.2110}} & \textbf{\textit{0.3906}} & 0.9875&\textbf{\textit{0.3112}}\\ 
        \cline{2-8}
        & VGP-7D   &\textbf{\textit{0.7736}} & 0.5062 & 0.2498 & 0.4206 & \textbf{\textit{0.9902}} & 0.3595\\\hline
        \multirow{2}{*}{Feb - 300} 
        & MWGP-S   & 0.9084 & 0.5857 & 0.2664 & 0.5127 & 0.9529&0.4568\\ 
        \cline{2-8} & VGP-5D & \textbf{\textit{0.8387}} & \textbf{\textit{0.5270}} & \textbf{\textit{0.2657}} & \textbf{\textit{0.4734}} & \textbf{\textit{0.9635}} & \textbf{\textit{0.4074}}\\\hline
        \multirow{3}{*}{Jan-Mar - 300} & MWGP-ST    & 0.8475 & 0.5522 & 0.2528 & 0.4726 & 0.9595&0.3634\\
        \cline{2-8}
        & KFD   & 0.8671 & 0.5032 & 0.2576 & 0.4568 & 0.9612 & \textbf{\textit{0.3575}}\\ 
        \cline{2-8}
        & VGP-7D   &\textbf{\textit{0.7693}} & \textbf{\textit{0.4885}} & \textbf{\textit{0.2417}} & \textbf{\textit{0.4206}} & \textbf{\textit{0.9693}} & 0.3641\\\hline
        \multirow{2}{*}{Feb - 1500}
        & MWGP-S   &\textbf{\textit{0.2901}} & 0.1859 & 0.0841 & 0.1668 & \textbf{\textit{0.9653}}&0.1393\\ 
        \cline{2-8} & VGP-5D & 0.2999 & \textbf{\textit{0.1556}} & \textbf{\textit{0.0740}} & \textbf{\textit{0.1575}} & 0.9627 & \textbf{\textit{0.1317}}\\\hline
        \multirow{3}{*}{Jan-Mar - 1500} & MWGP-ST   &0.2695 & 0.1661 & 0.0816 & 0.1553 & 0.9706&0.1146\\
        \cline{2-8}
        & KFD    & 0.2530 & 0.1559 & 0.0739 & 0.1385 & 0.9746&0.1233\\ 
        \cline{2-8}
        & VGP-7D   &\textbf{\textit{0.1938}} & \textbf{\textit{0.1367}} & \textbf{\textit{0.0636}} & \textbf{\textit{0.1170}} & \textbf{\textit{0.9844}} & \textbf{\textit{0.1013}}\\\hline
    \end{tabular}
    \label{tab:prediction_compare_regional2}
\end{table}

It is worth mentioning that the rankings based on point prediction accuracy do not always align with those incorporating both prediction and uncertainty quantification. For instance, at the 10 dbar pressure level in Region II, VGP achieved the lowest RMSE and the highest $R^2$, whereas KFD scored the best on other metrics, suggesting a misalignment between the squared loss and the negative log-likelihood.

\section{Conclusion}\label{Sec_Conclusion}
In this paper, we propose a unified Gaussian process (GP) regression framework for modeling Argo datasets, demonstrating that a generic nonparametric approach can outperform traditional, fine-tuned regression through data-driven training. Standard Argo modeling often relies on a two-stage estimation framework with prescribed mean functions and localized moving-window schemes. In contrast, our ``one-stop'' model eliminates the need for such manual tuning and complex parametric structures. To handle the cubic computational scaling of classical GPs ($O(n^3)$), we leverage the Vecchia approximation \citep{vecchia_estimation_1988, katzfuss_general_2021}, which reduces the complexity of the likelihood and gradient evaluations to quasi-linear in $n$. This scalability allows us to reduce reliance on localized modeling and enables a more holistic representation of global oceanographic variables. Our work shows that GPs can be effectively deployed on a massive scale to capture complex trends without sacrificing the rigorous probabilistic foundations of the model.

We evaluate our approach against two established benchmarks in the literature: the moving-window GP regression proposed by \citet{KuuselaStein18} and the spatio-temporal kriging framework for functional data by \citet{yargersh22}. The methods employ a two-stage estimation process, utilizing weighted least squares for prescribed mean structures followed by local GP regression on the resulting anomalies. While highly effective, such multi-stage procedures require careful calibration of local moving windows and parametric mean specifications, which can vary depending on the spatial coverage, temporal span, and depth of the dataset. Our unified GP framework offers an alternative by streamlining these requirements into a single, scalable modeling step. The results demonstrate that our ``one-stop'' approach achieves performance that compares favorably with the benchmark methods while providing a cohesive and interpretable tool for large-scale oceanographic applications.

A promising avenue for future research involves extending our unified framework to incorporate nonstationary covariance structures, enabling the joint modeling of Argo variables across the vertical dimension alongside horizontal space and time.

\section{Acknowledgment}
The authors acknowledge the support from the University of Houston, which provided the resources and environment that facilitated this research. We would also like to thank Drew Yarger for sharing resources that supported our replication of the method in \citet{yargersh22}. These Argo data were collected and made freely available by the International Argo Program and the national programs that contribute to it.  (\url{https://argo.ucsd.edu},  \url{https://www.ocean-ops.org}).  The Argo Program is part of the Global Ocean Observing System. 

\setcitestyle{numbers}
\bibliographystyle{apalike}
\bibliography{zotero}

\newpage

\section*{Appendix}

\begin{table}[ht]
    \centering
    \footnotesize
    \caption{Fitted covariance parameters in the one-stop GP model in the global experiment scenario.}
    \begin{tabular}{c|c|c|c|c|c|c|c|c|c|c}
    \hline
    Depth& Time & scale & \multicolumn{7}{|c|}{range} &  nugget\\
    \cline{3-11}
    level& window &$\sigma^2$ & $l$ & $L_s$ & $L_c$ & $p$ & $y$ & $d_s$ & $d_c$ &  $\tau^2$\\
    \hline

\multirow{2}{*}{10}&Feb&134.1736 & 4.5938 & 21.274 & 8.8343 & 532.8519 & 27.4576 & &&0.0021\\

\cline{2-11}&Jan-Mar&113.3547 & 3.706 & 22.2707 & 7.115 & 930.7529 & 18.408 & 272.6733 & 266.8972 & 0.0019\\
\hline
\multirow{2}{*}{300}&Feb&201.4941 & 5.178 & 37.9617 & 10.3863 & 484.3145 & 93.7521 &&& 0.002\\

\cline{2-11}&Jan-Mar&235.779 & 5.8282 & 51.4382 & 12.7972 & 777.9795 & 117.2012 & 644.8486 & 583.4145 & 0.0018\\
\hline
\multirow{2}{*}{1500}&Feb&6.0567 & 5.7418 & 19.9304 & 14.4351 & 167.289 & 228.8924 &&& 0.002\\

\cline{2-11}&Jan-Mar&7.3358 & 6.3486 & 13.4056 & 19.1399 & 344.1378 & 253.3443 & 386.3102 & 316.4217 & 0.0018\\
\hline
    \end{tabular}
    \label{tab:GP-parameters}
\end{table}

\begin{table}[ht]
    \centering
    \footnotesize
    \caption{Fitted covariance parameters in the one-stop GP model for the region centered at $(30^\circ \text{S}, 150^\circ \text{W})$.}
    \begin{tabular}{c|c|c|c|c|c|c|c|c|c|c}
    \hline
    Depth& Time & scale & \multicolumn{7}{|c|}{range} &  nugget\\
    \cline{3-11}
    level& window &$\sigma^2$ & $l$ & $L_s$ & $L_c$ & $p$ & $y$ & $d_s$ & $d_c$ &  $\tau^2$\\
    \hline

\multirow{2}{*}{10}&Feb&88.4731 & 10.8673 & 32.4555 & 73.4795 & 813.6854 & 16.0166 &&& 0.0022\\

\cline{2-11}&Jan-Mar&58.1434 & 7.7213 & 21.4388 & 42.5546 & 1025.2811 & 10.0141 & 38.164 & 89.8908 & 0.0018\\
\hline
\multirow{2}{*}{300}&Feb&92.7444 & 8.3656 & 51.2693 & 38.0606 & 821.6799 & 71.7433 &&& 0.0033\\

\cline{2-11}&Jan-Mar&93.2071 & 6.6363 & 27.7002 & 30.8697 & 760.2389 & 43.2671 & 601.3635 & 847.9366 & 0.0024\\
\hline
\multirow{2}{*}{1500}&Feb&0.6266 & 13.3205 & 195.5327 & 242.1713 & 223.0109 & 248.0358 &&& 0.0259\\

\cline{2-11}&Jan-Mar&0.5417 & 14.2501 & 212.053 & 199.6611 & 261.412 & 274.0756 & 261.5292 & 257.598 & 0.025\\
\hline
    \end{tabular}
    \label{tab:GP-parameters_30150}
\end{table}

\begin{table}[ht]
    \centering
    \footnotesize
    \caption{Fitted covariance parameters in the one-stop GP model for the region centered at $(30^\circ \text{N}, 30^\circ \text{W})$.}
    \begin{tabular}{c|c|c|c|c|c|c|c|c|c|c}
    \hline
    Depth& Time & scale & \multicolumn{7}{|c|}{range} &  nugget\\
    \cline{3-11}
    level& window &$\sigma^2$ & $l$ & $L_s$ & $L_c$ & $p$ & $y$ & $d_s$ & $d_c$ &  $\tau^2$\\
    \hline

\multirow{2}{*}{10}&Feb&108.4705 & 3.3366 & 365.2706 & 6.5311 & 453.7382 & 15.7292 &&& 0.0018\\

\cline{2-11}&Jan-Mar&141.3146 & 4.1583 & 523.0554 & 10.4651 & 1293.0526 & 17.3872 & 105.5127 & 421.9463 & 0.0019\\
\hline
\multirow{2}{*}{300}&Feb&163.5792 & 4.463 & 396.9711 & 8.5982 & 720.2092 & 21.8352 &&& 0.0018\\

\cline{2-11}&Jan-Mar&190.7658 & 5.0072 & 117.3103 & 9.326 & 1444.1211 & 174.0469 & 335.1567 & 499.361 & 0.0019\\
\hline
\multirow{2}{*}{1500}&Feb&13.0906 & 8.8217 & 9.173 & 50.2513 & 187.8329 & 42.289 &&& 0.0017\\

\cline{2-11}&Jan-Mar&16.062 & 7.9263 & 10.3249 & 99.5459 & 442.2838 & 83.7797 & 998.5292 & 763.4305 & 0.0018\\
\hline
    \end{tabular}
    \label{tab:GP-parameters_3030}
\end{table}

\begin{figure}[htbp]
    \centering
    \caption{Prediction residuals at 10 dbar with models trained on three-month floats.}
    \label{fig:Res_10}
    \subfloat[KFD]{
    \includegraphics[width=0.485\linewidth]{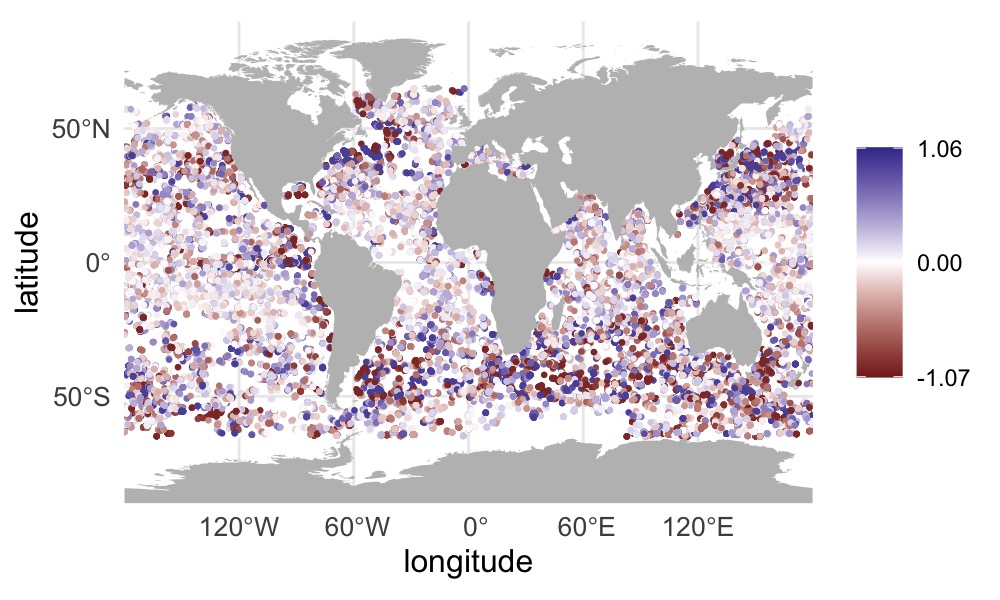}
    \label{fig:Res_10_KFD}
    }
    \hfill
    \subfloat[VGP-7D]{
    \includegraphics[width=0.485\linewidth]{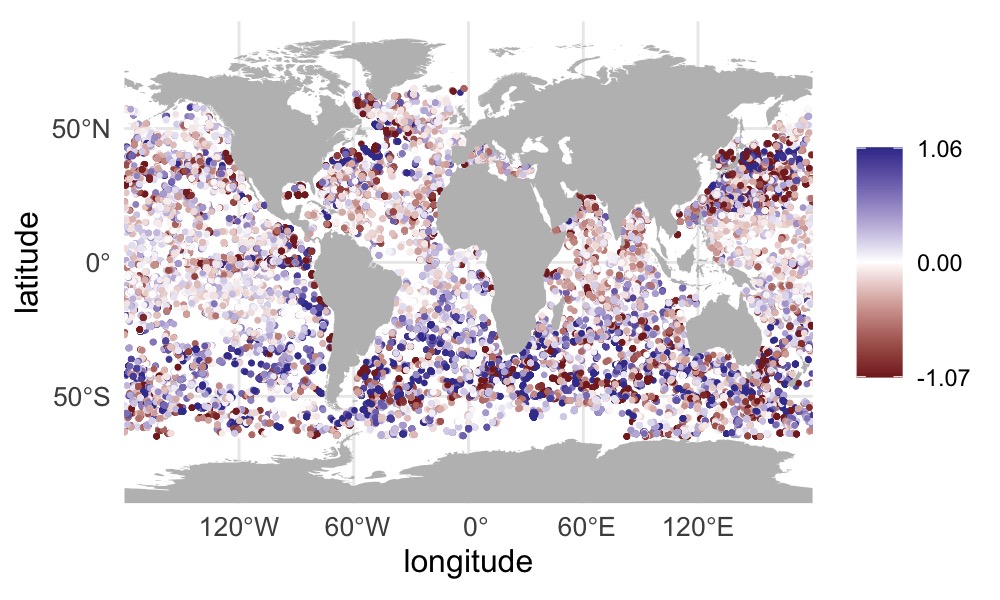}
    \label{fig:Res_10_VGP-7D}
    }\\
    \subfloat[MWGP-ST]{
    \includegraphics[width=0.485\linewidth]{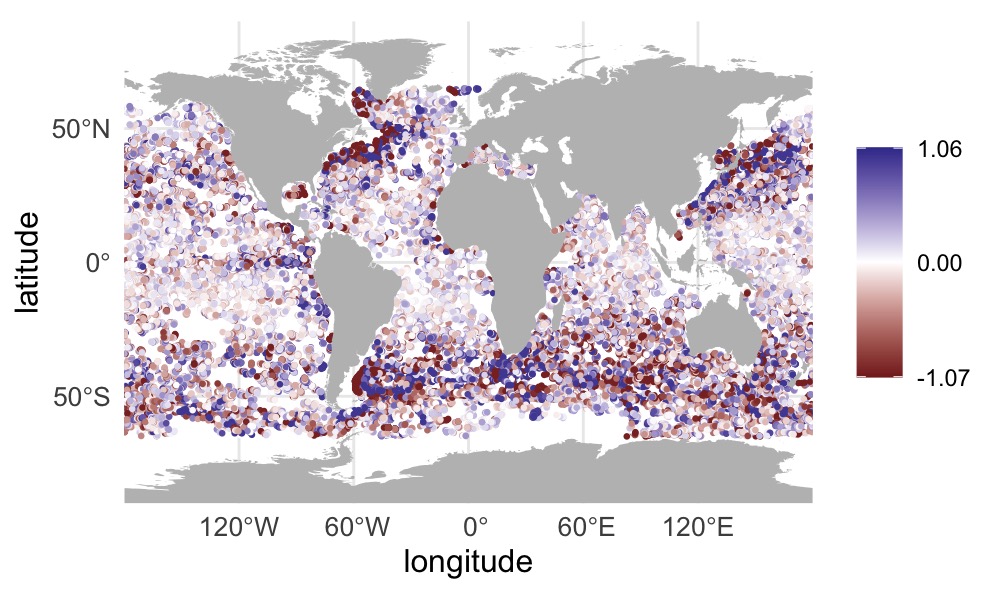}
    \label{fig:Res_10_MWGP}
    }
\end{figure}

\begin{figure}[htbp]
    \centering
    \caption{Prediction residuals at 300 dbar with models trained on three-month floats.}
    \label{fig:Res_300}
    \subfloat[KFD]{
    \includegraphics[width=0.485\linewidth]{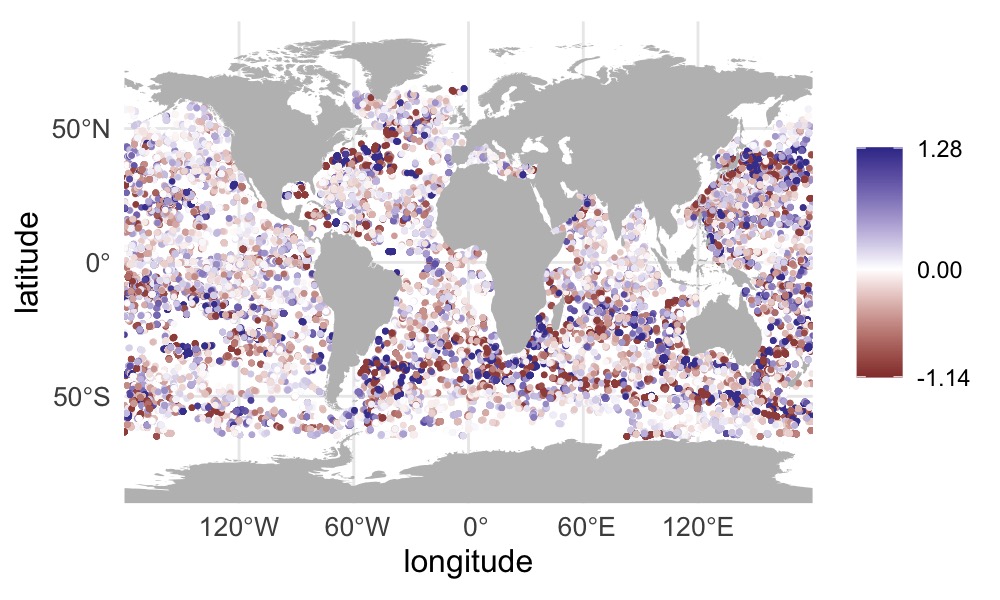}
    \label{fig:Res_300_KFD}
    }
    \hfill
    \subfloat[VGP-7D]{
    \includegraphics[width=0.485\linewidth]{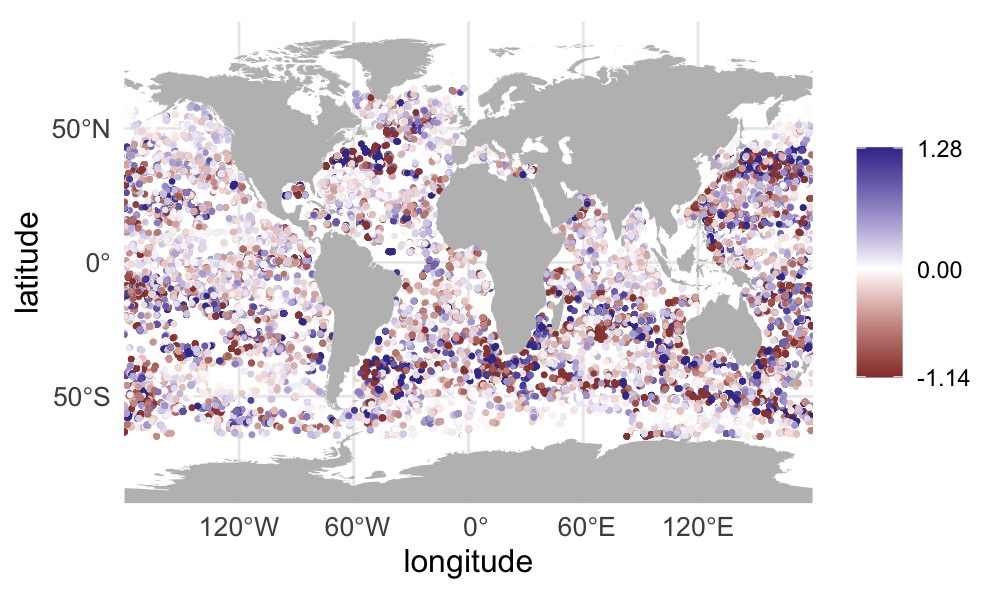}
    \label{fig:Res_300_VGP-7D}
    }\\
    \subfloat[MWGP-ST]{
    \includegraphics[width=0.485\linewidth]{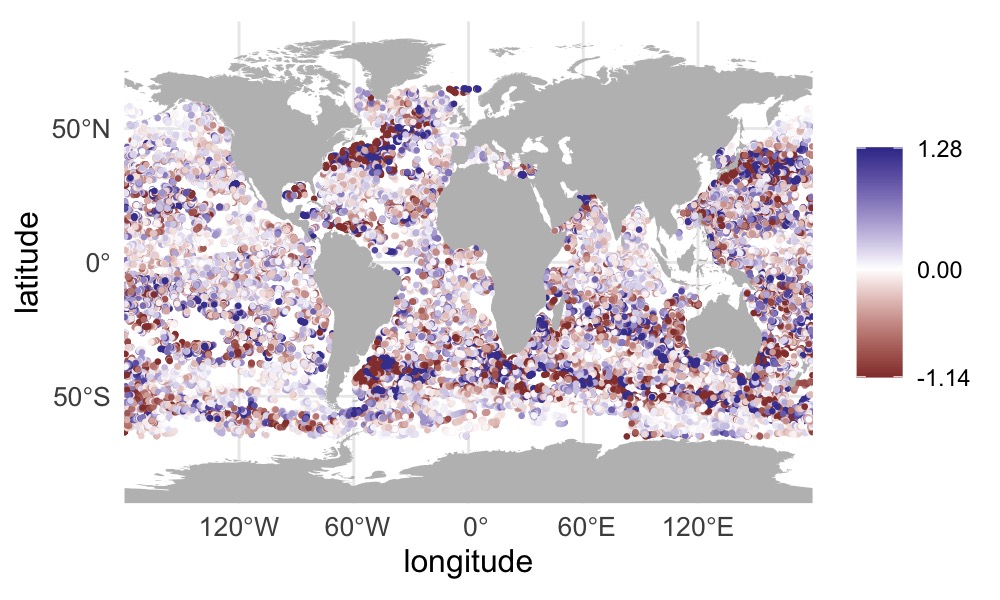}
    \label{fig:Res_300_MWGP}
    }
\end{figure}
\end{document}